\documentclass[two column,english,prl,superscriptaddress,preprintnumbers,amsmath,amssymb,floatfix]{revtex4-1}
\pdfpageattr {/Group << /S /Transparency /I true /CS /DeviceRGB>>}
\usepackage[T1]{fontenc}
\usepackage[latin9]{inputenc}
\usepackage{dcolumn}
\usepackage{bm}
\usepackage{graphicx}
\usepackage{color}
\usepackage{esint}
\usepackage{babel}
\usepackage{amsfonts}
\usepackage{slashed}
\usepackage{enumerate}

\begin{document}


\title{Universality and scaling in a charge two-channel Kondo device}

\author{A. K. Mitchell}
\affiliation{Institute for Theoretical Physics, Utrecht University, Leuvenlaan 4, 3584 CE Utrecht, Netherlands}

\author{L. A. Landau}
\affiliation{Raymond and Beverly Sackler School of Physics and Astronomy, Tel-Aviv University, Tel Aviv 69978, Israel}

\author{L. Fritz}
\affiliation{Institute for Theoretical Physics, Utrecht University, Leuvenlaan 4, 3584 CE Utrecht, Netherlands}

\author{E. Sela}
\affiliation{Raymond and Beverly Sackler School of Physics and Astronomy, Tel-Aviv University, Tel Aviv 69978, Israel}


\begin{abstract}
\noindent We study a charge two-channel Kondo model, demonstrating that recent experiments [Iftikhar \emph{et al}, Nature \textbf{526}, 233 (2015)] realize an essentially perfect quantum simulation -- not just of its universal physics, but also \emph{nonuniversal} effects away from the scaling limit. Numerical renormalization group calculations yield conductance lineshapes encoding RG flow to a critical point involving a free Majorana fermion. By mimicking the experimental protocol, the experimental curve is reproduced quantitatively
over 9 orders of magnitude, 
although we show that far greater bandwidth/temperature separation is required to obtain the universal result. 
Fermi liquid instabilities are also studied: In particular, our exact analytic results for non-linear conductance provide predictions \emph{away from thermal equilibrium}, in the regime of existing experiments.
\end{abstract}

\maketitle


\textit{Introduction and results.--} 
The Kondo effect provides a paradigmatic example for universality: systems with very different microscopic details exhibit the same behavior in terms of rescaled temperature $T/T_K$, where the Kondo temperature $T_K$ is an emergent low-energy scale characteristic of the particular system \cite{Hewson}. Scaling collapse of data to a single universal curve has been demonstrated in diverse contexts, ranging from magnetic impurities in metals \cite{Hewson}, semiconductor \cite{DGG,*WdW,*Cronenwett} or nanotube \cite{nygaard} quantum dots, and molecules in break junctions \cite{kondo_molec}.
Although simplified, the Kondo model, comprising a single spin-$\tfrac{1}{2}$ quantum `impurity' exchange-coupled to a single conduction electron channel, describes accurately the low-energy physics of many such systems due to the single parameter scaling -- provided $T_K$ is much smaller than bare energy scales, such as the conduction electron bandwidth $D$.

In the two-channel Kondo (2CK) effect \cite{Nozieres80}, two spinful channels compete to screen a single impurity, leading to a frustrated non-Fermi liquid (NFL) quantum critical point involving a free Majorana fermion \cite{EK}.
At the critical point, there is again universality in terms of $T/T_K$ when $T_K/D\ll 1$. 
However, on detuning away from the critical point, the frustration is relieved by a Fermi liquid (FL) instability, which drives the system toward an FL ground state on the scale $T^{*}$. Provided $T^{*} \ll T_K \ll D$, there are two successive crossovers on reducing temperature: first to the NFL critical point, and then away from it \cite{Glaz,Sela11,*mitchell2012universal}. The FL crossover is then a universal function of $T/T^{*}$. 

While these issues of universality and quantum criticality are being explored in bulk strongly correlated materials (e.g.\ high-$T_c$ superconductors or heavy fermion systems) \cite{subir}, the 2CK effect can be realized in highly controllable and tuneable semiconductor quantum dot devices~\cite{Potok07,Keller15,Iftikhar15}.
The experimental observation of critical NFL physics in such devices depends on minimizing $T^{*}$, and exploring the regime $T\gg T^{*}$. In particular, the measured conductance encodes a nontrivial renormalization group (RG) flow on reducing temperature.

\begin{figure}[b]
  \centering
\includegraphics[width=7cm]{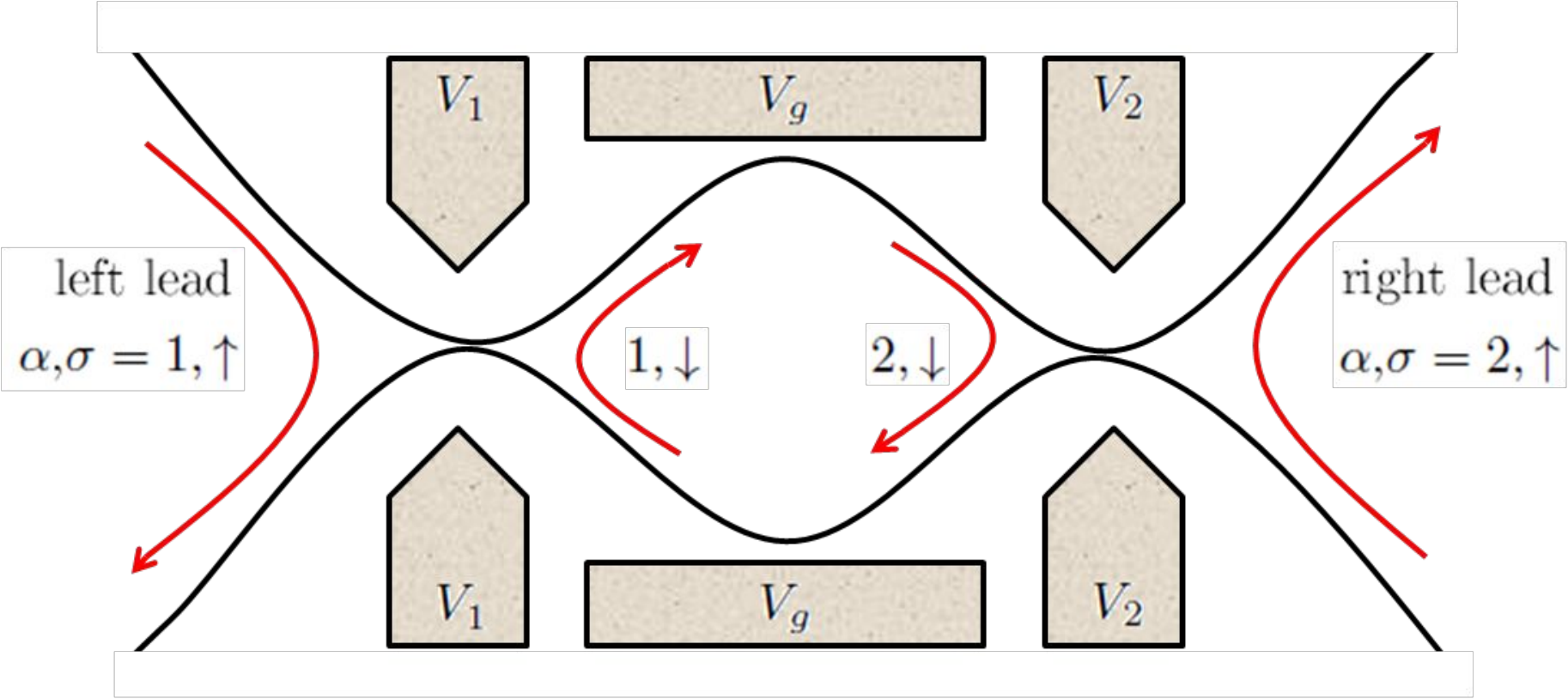}
  \caption{Schematic of the charge 2CK device: Two QPCs, with transmissions $\tau_{1,2}$ controlled by $V_{1,2}$, connect the quantum dot to two leads. Gate voltage $V_g$ controls the dot charge.}
  \label{fig:model}
\end{figure}

A novel \emph{charge} 2CK device \cite{Furusaki95} was studied recently in the remarkable experiments of Ref.~\cite{Iftikhar15} (see Fig.~\ref{fig:model}). 
The results shed new light on scaling and universality in the 2CK effect from an unusual perspective: the electronic spin degrees of freedom are quenched by a field, while two nearly degenerate macroscopic charge states of a large quantum dot play the role of a pseudospin \cite{matveev}. The dot is connected to two metallic leads via quantum point contacts (QPCs), with transmissions $0\le \tau_{1,2} \le 1$. Pseudospin flips are caused by tunneling through the QPCs; the Kondo exchange is therefore \emph{first-order} in the tunneling, and becomes large when the QPCs are opened (by contrast, spin flips are virtual second-order processes in conventional spin-Kondo devices).  Conductance across the dot, due to a bias voltage between the leads, involves successively flipping the pseudospin by tunneling at one QPC, and then flipping it back by tunneling at the other. The finite dot capacitance prevents charge buildup, suppressing transitions out of the pseudospin-$\tfrac{1}{2}$ manifold and strongly correlating  tunneling events. 

The experiment provides compelling evidence for 2CK RG flow: tuning $\tau_1\approx \tau_2$, the measured conductance and effective transmissions increase on decreasing temperature. The low-temperature proliferation of pseudospin flips driving this conductance enhancement is the classic hallmark of Kondo physics. At large transmission $\tau_{1,2}\sim 1$, the conductance approaches $G=e^2/(2h)$, the limit for two quantum resistors in series. Furthermore, the competition between screening channels that characterizes the 2CK effect is observed in the experiment upon detuning $\tau_1 \ne \tau_2$, whence the frustration of the 2CK state is relieved and the conductance drops again. 

\begin{figure}[t]
  \centering
\includegraphics[width=7cm]{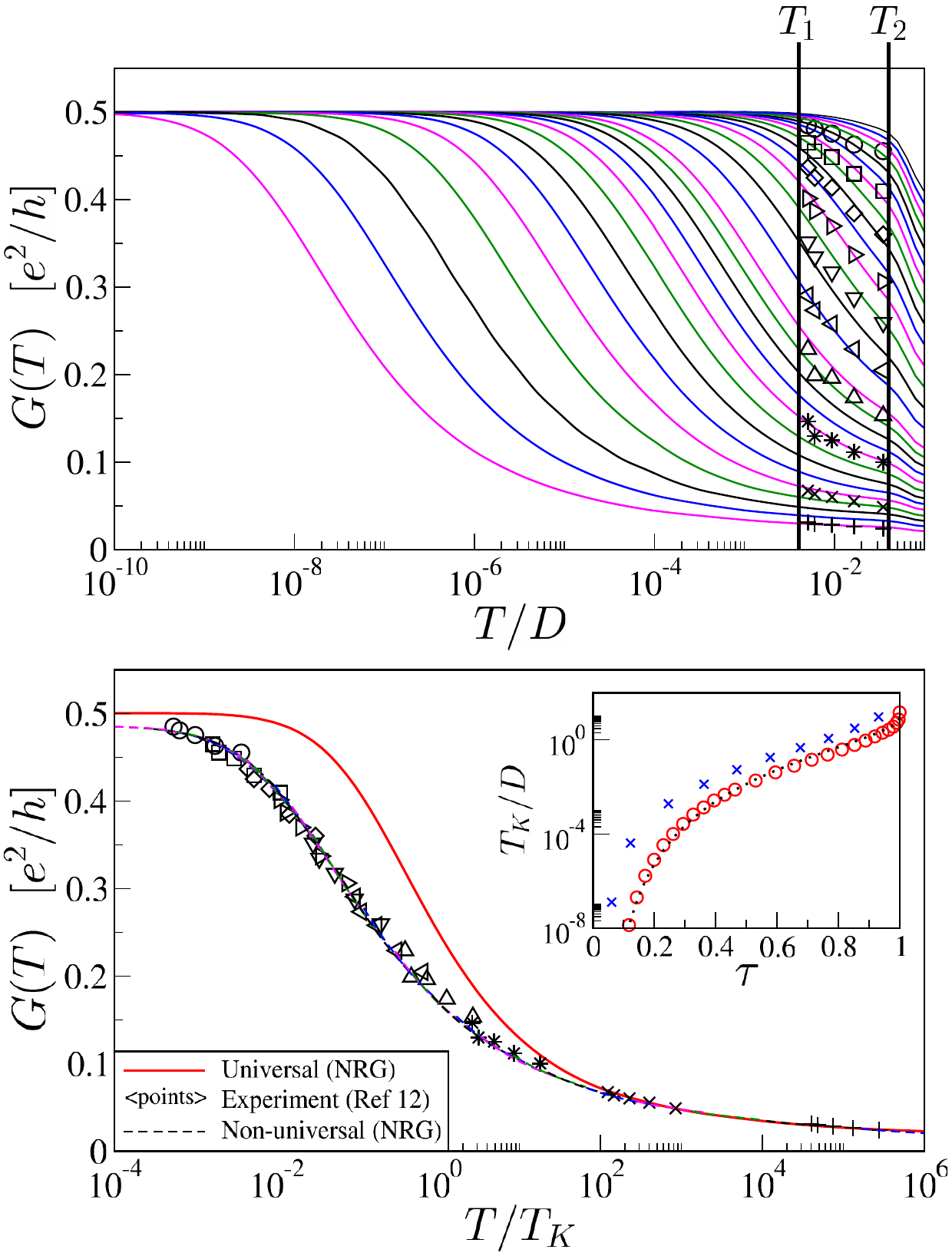}
  \caption{Conductance $G(T)$ of the charge 2CK device along the NFL crossover. \emph{Upper panel:} NRG results (lines) for different transmissions from $\tau_1$=$\tau_2\simeq 0.1$ to $1$ ($J/D$=$0.1$ to $0.6$) as an entire function of temperature $T/D$. Points are the experimental data of Ref.~\cite{Iftikhar15}, fit using $D$=$2.3$K. \emph{Lower:} Universal NFL crossover $G(T/T_K)$ from NRG in the scaling limit (solid line), compared with rescaled experimental data (points) and rescaled NRG data (dashed lines) from the temperature window $T_1$<$T$<$T_2$, with $T_2$=$10T_1$=$0.04D$. Both curves have the same asymptotes $G(T)\sim \ln^{-2}(T/T_K)$ for $T\gg T_K$ and $G(0)-G(T)\sim (T/T_K)$ for $T\ll T_K$. \emph{Inset:} $T_K$ vs $\tau$ for NRG (circles) and experiment (crosses). Dotted line is $T_K\sim D \nu J \exp[-\pi/4\nu J]$, with $\nu J$ related to $\tau$ via Eq.~\ref{eq:tau}.
	\label{fig:NFL}} 
\end{figure}

A key result of Ref.~\cite{Iftikhar15} is the \emph{seemingly} universal NFL conductance crossover [which we denote $\tilde{G}(T/T_K)$], which spans an unprecedented range of $T/T_K$ over 9 orders of magnitude. Although the experiment is confined to a narrow temperature range 11.5--80 mK, the entire curve is uncovered by patching together data for different systems, each with a different transmission and hence $T_K$. The wide range of $T/T_K$ follows from the exponential dependence of the Kondo temperature on transmission. The reconstruction of $\tilde{G}(T/T_K)$ is shown in Fig.~\ref{fig:NFL}.

This approach assumes single-parameter scaling of the conductance, $G(T/T_K)$. However, this universality arises strictly only for $T_K/D \rightarrow 0$ (the `scaling limit'). In general, the conductance $G(T/T_K,T_K/D)$ is a function of both $T/T_K$ and $T_K/D$. 
In this paper, we show that this has significant consequences for the interpretation of experimental results: data collected at large transmission (i.e.\ large $T_K$) are not in the universal regime, and belong to a nonuniversal cut through the surface of $G(T/T_K,T_K/D)$. We show that the experimental curve $\tilde{G}(T/T_K)$ does not match the true universal curve $G(T/T_K)$, obtained by solving the charge 2CK model using the numerical renormalization group (NRG) \cite{bulla2008numerical,*iNRG,*iNRG2,Lebanon03,*Lebanon2}.

However, by mimicking the experimental protocol, we \emph{quantitatively} reproduce the entire experimental curve -- see lower panel of Fig~\ref{fig:NFL}.\ Not only does this substantiate the above scaling arguments, but it also demonstrates that the experimental system of Ref~\cite{Iftikhar15} is an essentially perfect quantum simulation of the charge 2CK model, describing faithfully not just its asymptotic universal physics (which is blind to microscopic details), but including also nonuniversal effects of the finite bandwidth.

\begin{figure}[t]
  \centering
\includegraphics[width=7cm]{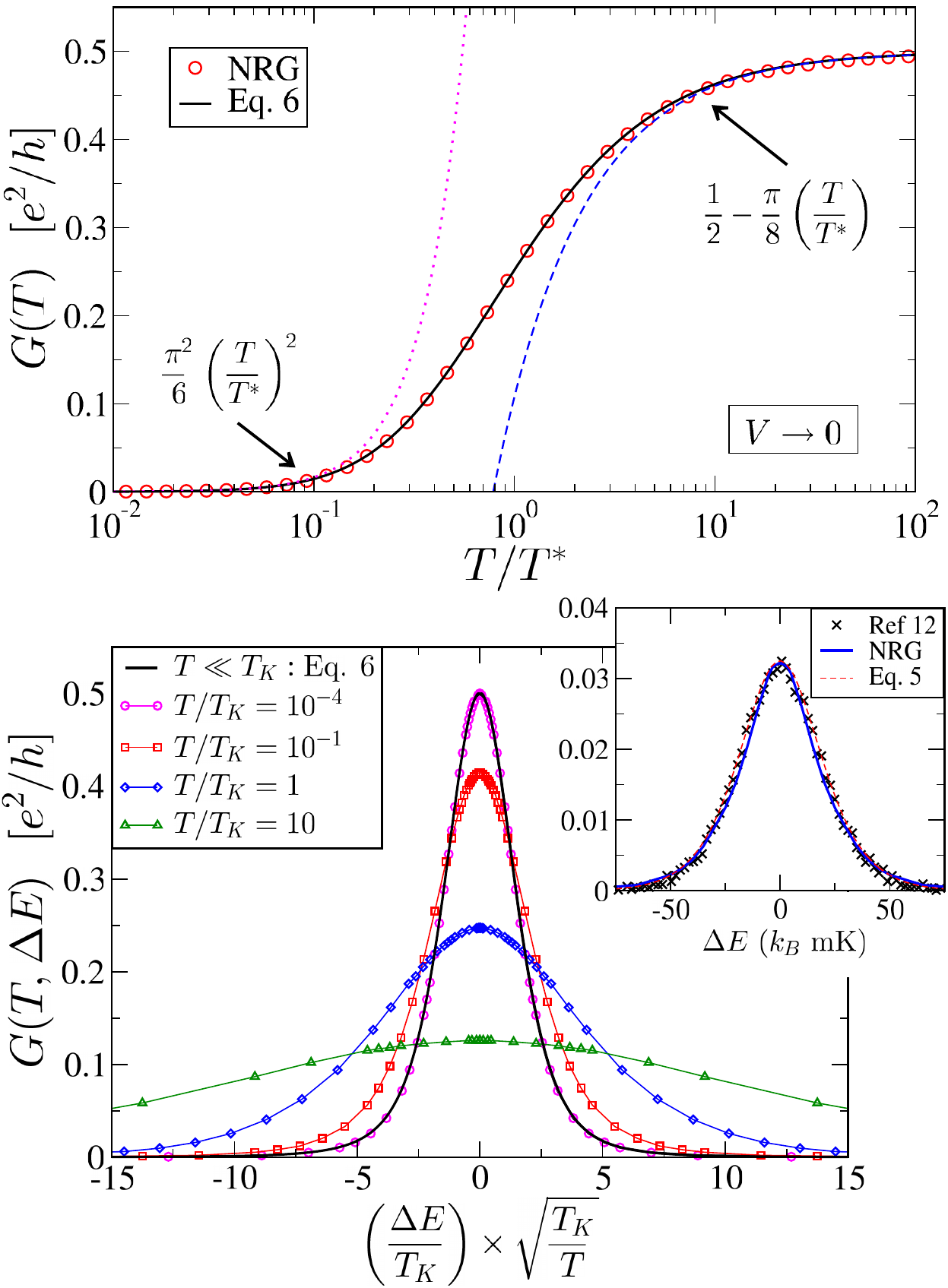}
  \caption{\emph{Upper panel}: Low temperature universal FL conductance crossover as a function of $T/T^{*}$ (valid when $T^{*}\ll T_K$). Points are NRG data; solid line is the exact analytic result, Eq.~\ref{eq:FL}. \emph{Lower panel}: Coulomb blockade peaks at various temperatures, obtained from NRG. The universal result is shown as the black line. \emph{Inset}: Experimental data at weak transmission $\tau=0.06$ from Ref.~\cite{Iftikhar15} (points), compared with NRG fit (solid line) and Eq.~\ref{eq:classical} (dashed line).
	}
  \label{fig:FL}
\end{figure}

As with the experiment, we also study the FL crossover, generated either by transmission asymmetry $\tau_1 \ne \tau_2$, or by moving away from the dot charge degeneracy point -- see Fig.~\ref{fig:FL}. The conductance $G(T/T_K,T/T^{*},T_K/D)$ is now also a function of $T/T^{*}$. Even in the scaling limit $T_K\ll D$, there is now a \emph{family} of universal curves for different ratios of $T^{*}/T_K$, which we explore with NRG. Agreement is found with experiment at weak transmission where universality is expected. 

\begin{figure}[t]
  \centering
\includegraphics[width=7.5cm]{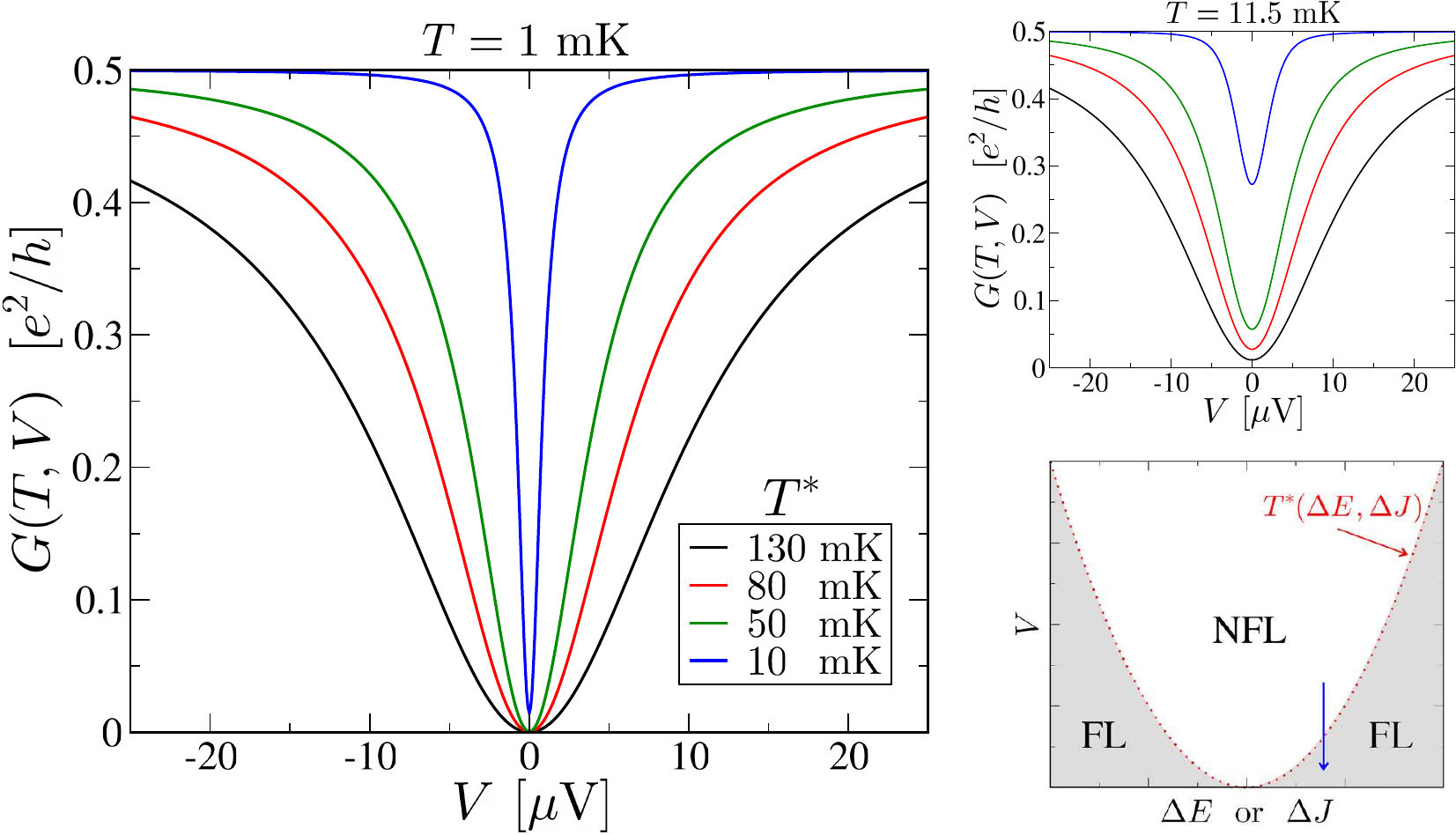}
  \caption{Nonequilibrium critical behavior. \emph{Main figure}: Nonlinear conductance at $T$=$1$mK, for various $T^{*}$. \emph{Top right}: Same but for experimental base temperature $T$=$11.5$mK. \emph{Bottom right}: FL crossover (blue arrow) controlled by $eV/T^{*}$.}
  \label{fig:noneq}
\end{figure}

Furthermore, in the case of good scale separation $T^{*}\ll T_K$, the FL crossover exhibits single-parameter scaling in $T/T^{*}$. Here we use Abelian bosonization methods \cite{EK} to obtain an exact analytic result for $G(T/T^{*},eV/T^{*})$, which remarkably also describes the \emph{nonequilibrium} conductance at finite bias $V$. We show that this result does not require $T_K\ll D$, meaning that the same lineshape is obtained at large transmission. Therefore, the physics of the charge 2CK model away from thermal equilibrium can be explored in the parameter regime of existing experiments, at realistic base temperatures -- see Fig.~\ref{fig:noneq}.


\textit{Model and Experiment.--}
The experiment Ref.~\onlinecite{Iftikhar15} consists of a large metallic quantum dot connected via single-channel quantum point contacts to two leads in the quantum Hall regime (Fig.~\ref{fig:model}). The effective Hamiltonian is
\begin{eqnarray}
\label{eq:H_c2ck}
\begin{aligned}
H_{\text{K}}=\sum_{\alpha=1,2} \Big [J_{\alpha} &\sum_{k,k'}\left ( c_{\alpha \uparrow k}^{\dagger}c_{\alpha \downarrow k'}^{\phantom{\dagger}} \hat{S}^{-} + c_{\alpha \downarrow k}^{\dagger}c_{\alpha \uparrow k'}^{\phantom{\dagger}} \hat{S}^{+}\right ) \\
+&\sum_{k,\sigma} \epsilon_{\alpha \sigma k}^{\phantom{\dagger}} c_{\alpha\sigma k}^{\dagger}c_{\alpha \sigma k}^{\phantom{\dagger}}   \Big ]  +\Delta E \hat{S}^z \;, 
\end{aligned}
\end{eqnarray}
where $\sigma=\uparrow$ denotes spinless electrons in the two leads (labelled $\alpha=1,2$), while $\sigma=\downarrow$ denotes spinless electrons on the dot. The two QPCs are described
by \emph{independent} electronic systems labelled $\alpha=1,2$. Neglecting coherent transport of electrons between the QPCs is justified \cite{averin1989macroscopic,*PhysRevLett.65.2446} for large quantum dots with level spacing $\delta \ll T$ (in Ref.~\onlinecite{Iftikhar15}, $\delta \sim  0.2 \mu$K < $10^{-4} T$).

The total number of dot electrons $N_{\downarrow}=\sum_{\alpha} N_{\alpha\downarrow}$ \cite{note:numop} is changed when an electron tunnels through either QPC, such that $\hat{S}^{\pm}|N_{\downarrow}\rangle = |N_{\downarrow}\pm 1\rangle$. However, the dot is in the Coulomb blockade regime, with charging energy $E_c \cong 290$ mK. 
By tuning the gate voltage $V_g\rightarrow V_g^0$ to the edge of a Coulomb blockade step, dot states with $N_{\downarrow}=N_0$ and $N_0+1$ electrons are degenerate, while $N_{\downarrow}<N_0$ and $N_{\downarrow}>N_0+1$ are much higher in energy and essentially inaccessible at experimentally relevant temperatures $T\ll E_c$  (our results are unchanged by including more charge states). The two degenerate charge states form a pseudospin-$\tfrac{1}{2}$, with $\hat{S}^z = \frac{1}{2} \left( |N_0+1 \rangle \langle N_0+1 |-|N_0 \rangle \langle N_0 | \right)$. 
Detuning the gate splits the states by an energy $\Delta E\propto (V_g-V_g^0)$, equivalent to the pseudospin field $\Delta E \hat{S}_z$ in Eq.~\ref{eq:H_c2ck}. 

Eq.~\ref{eq:H_c2ck} is a spin anisotropic 2CK model, with two electronic channels $\alpha=1,2$ defined around the left and right QPCs, and two macroscopic dot charge states playing the role of the `impurity' spin-$\tfrac{1}{2}$. The QPC transmissions $\tau_{\alpha}$ determine the spin-flip Kondo couplings $J_{\alpha}$ via \cite{Lebanon03,*Lebanon2},
\begin{eqnarray}
\label{eq:tau}
\tau_\alpha = 4 (\pi \nu_{\alpha} J_{\alpha})^2/[1+(\pi \nu_{\alpha} J_{\alpha})^2]^2 \;,
\end{eqnarray}
where the effective density of states $\nu_{\alpha}=\sqrt{\nu_{\alpha\uparrow} \nu_{\alpha\downarrow}}$ relevant for spin flips at QPC $\alpha$ depends on both the lead ($\sigma=\uparrow$) and dot ($\sigma=\downarrow$) densities. Although $\nu_{\alpha\sigma}$ will depend on $\alpha$ and $\sigma$ in a real device, $\nu_{\alpha}J_{\alpha}$ depends only on $\tau_{\alpha}$ via Eq.~\ref{eq:tau}, and can be manipulated by gate $V_{\alpha}$ in experiment. $\nu_{\alpha}$ itself is immaterial since the Kondo physics is controlled by $\nu_{\alpha}J_{\alpha}$. 
Since the leads comprise chiral 1D fermions in quantum Hall edge states \cite{Iftikhar15}, the role of interactions \cite{HJ_c2ck,*kakashvili2007enhanced} is suppressed, and the low-energy density of state is constant \cite{note:dos}; for simplicity we take $\nu_{\alpha\sigma}(\omega)\propto \sum_k \delta(\epsilon_{\alpha \sigma k} - \omega) = \nu \Theta(D-|\omega|)$ with $\nu=1/(2D)$ and $D$ the conduction electron bandwidth.


\textit{Conductance.--}
Our primary interest is in the experimentally observable serial differential conductance across the dot, due to a voltage difference between leads 1 and 2 (i.e.\ $\sigma=\uparrow$ electrons only). The geometry does not allow a simple formulation in terms of Green's functions, as for standard (single channel) devices \cite{MW}. The ac conductance $G(\omega,T)$ at zero-bias $V\rightarrow 0$ can however be obtained within linear response from the Kubo formula~\cite{izumida1997many},
\begin{equation}
\label{eq:cond_kubo}
G(\omega,T) = \left ( \frac{e^2}{h}\right ) \times \left [ \frac{2\pi \hbar^2 ~\text{Im}K(\omega, T)}{\hbar \omega} \right ] \;,
\end{equation}
where $K(\omega,T)$ is the Fourier transform of $K(t,T) = i\Theta(t) \langle [ \hat{\Omega}(t), \hat{\Omega}(0) ] \rangle_{T}$,
with $\hat{\Omega}=\tfrac{1}{2}(\dot{N}_{1\uparrow}-\dot{N}_{2\uparrow}) = \tfrac{1}{2}\sum_{k,k'}[( J_{1}^{\phantom{\dagger}}
c_{1 \uparrow k}^{\dagger}c_{1 \downarrow k'}^{\phantom{\dagger}} - J_{2 }^{\phantom{\dagger}}
c_{2 \uparrow k}^{\dagger}c_{2 \downarrow k'}^{\phantom{\dagger}})\hat{S}^{-} - \text{H.c.}]$. Here $K(\omega,T)$ is calculated numerically-exactly on the real axis using full-density-matrix NRG \cite{weichselbaum2007sum,note:nrg}. The desired dc conductance is obtained as $G(T)=G(\omega\rightarrow 0,T)$.

Fig.~\ref{fig:NFL} (upper panel) shows the calculated conductance $G(T)$ vs $T/D$ along the NFL crossover (lines) for systems with different transmissions ranging from $\tau_1=\tau_2\equiv \tau\simeq 0.1$ to $1$, yielding Kondo temperatures $T_K/D$ in the range $10^{-8}$ to $>1$ ($T_K$ is defined as the half-width at half-maximum, $G(T_K)=\tfrac{1}{4}$, as in the experiment). The sequence of lines show the onset of nonuniversal effects at higher energies, which become very pronounced at large transmission where the scaling condition $T_K\ll D$ is strongly violated. Also shown are the experimental data of Ref.~\cite{Iftikhar15} (points) fit using $D$=$2.3$K. 

Fig.~\ref{fig:NFL} (lower panel) shows the experimental data now rescaled in terms of $T/T_K$, precisely as in Ref.~\cite{Iftikhar15}, compared with rescaled NRG data (dashed lines) taken from the upper panel in the temperature window $T_1$<$T$<$T_2$, mimicking the experimental protocol. The evolution of $T_K$ is plotted in the inset. We find excellent agreement between the theory and experimental curves. However, they are rather different from the true universal curve $G(T/T_K)$, obtained by NRG in the scaling limit (in practice $\tau$=$0.07$).
We emphasize that the discrepancy between the experimental and universal theory curves is {\it neither} a shortcoming of the model or theoretical analysis {\it nor} the experiment, but is fully consistent once non-universal effects of finite bandwidth are accounted for.

The universal curve obtained at small transmission is independent of the specific value of $\tau$ (or equivalently $\nu J$). This universality goes beyond the specific model: For example, we also computed the conductance in a generalized model \cite{Lebanon03,*Lebanon2} including up to 20 higher energy charge states, in addition to the degenerate pair of ground states, which takes into account the finite charging energy term $4 E_c \hat{S}_z^2$. In this case we find that $T_K$ is suppressed at small $E_c$, since $E_c$ sets the effective bandwidth below which the two-level impurity model is valid. However, the universal crossover function $G(T/T_K)$ obtained by NRG is unchanged in the scaling regime $T_K \ll D,E_c$.

By contrast, conductance lineshapes obtained at large transmission do not exhibit universal scaling. Experimental data at $\tau$=$0.93$ with $T_K$ on the order of the bandwidth (set by $E_c$) belong to a nonuniversal curve.


\textit{Fermi Liquid Instability.--} 
Inherent to the 2CK physics is its quantum critical structure. The NFL state is delicate and unstable to various symmetry breaking perturbations \cite{affleck2CK}, including channel anisotropy $\Delta J$=$J_{1}$-$J_{2}$ (controlled by gates $V_1$ and $V_2$), and deviations from the charge degeneracy point, $\Delta E$ (controlled by $V_g$). These perturbations generate a new energy scale $T^{*}$, controlling the crossover to the FL ground state \cite{Glaz,note:matveev}. For small perturbations $\Delta J$, $\Delta E < T_K$, then $T^* \ll T_K$ viz:
\begin{eqnarray}
\label{eq:Ts}
T^{*} = c_1 T_K(\nu \Delta J)^2 + c_2(\Delta E)^2/T_K \;.
\end{eqnarray}
In this case the FL crossover is a universal function of $T/T^{*}$ only, controlled by leading dimension 1/2 operators. Symmetry breaking perturbations grow under RG to reduce the conductance along the FL crossover.

Fig.~\ref{fig:FL} (upper panel) shows the universal NRG result (points) for the FL conductance crossover, $G(T/T^{*})$ [see also analytic form below]. We have confirmed that the same crossover is obtained for $\Delta J$ and $\Delta E$ (or any combination of the two); only the emergent scale $T^{*}$ enters.
The universality of $G(T/T^{*})$, together with Eq.~\ref{eq:Ts}, can be used to make experimental scaling predictions. For example, conductance data should exhibit scaling collapse in terms of $(\Delta E/T_K)\times\sqrt{T_K/T} \propto (T/T^{*})^{-1/2}$ at low temperatures. NRG results in Fig.~\ref{fig:FL} (lower panel) show the $\Delta E$ dependence of the conductance as $T$ is decreased. We find that in practice the universal result (black line) is essentially recovered for $T/T_K\sim 10^{-4}$.

On the other hand, the experiments at small transmission $\tau=0.06$ were performed in the regime $T\gg T_K$. Here the classical sequential tunneling result should hold,
\begin{eqnarray}
\label{eq:classical}
G(T) \overset{T\gg T_K}{=} \tfrac{e^2/h}{\tau_1^{-1}+\tau_2^{-1}}\times \tfrac{\Delta E/T}{ \sinh(\Delta E/T)} \;,
\end{eqnarray}
with Kondo-renormalized transmissions, as also confirmed by the fit to NRG in the inset of Fig.~\ref{fig:FL}.

Focusing now on the universal regime $T^{*}\ll T_K$, we can also obtain exact analytic results for the FL crossover using Abelian bosonization methods \cite{EK}, exploiting the effective free fermion structure of the relevant dimension 1/2 operators. Refs.~\cite{Sela11,*mitchell2012universal} used such methods to make predictions for the FL crossovers occurring in a spin 2CK device (with different device geometry), which were recently confirmed in the experiments of Ref.~\cite{Keller15}.
For the charge 2CK device geometry, we adapt instead the methods of Schiller and Hershfield~\cite{Schiller98} (see also \cite{SelaAffleck,*sela2009nonequilibrium}). Remarkably, the \emph{nonequilibrium} crossover at finite bias voltage $V$ can be obtained, which holds for arbitrary temperatures (provided $eV,T\ll T_K$). Our exact result~is
\begin{eqnarray}
\label{eq:FL}
G(T,V) =\frac{e^2}{2h}\left[1 - \frac{T^*}{2 \pi  T} {\rm{Re}}\psi^{(1)} \left(\frac{1}{2}+\frac{T^*+i e V}{2 \pi  T} \right)\right],\;\;\;
\end{eqnarray}
where $\psi^{(1)} = \partial_z \psi(z)$ is the trigamma function. The zero-bias limit $V\rightarrow 0$ coincides with the result obtained starting from a completely different state of perfect transmission \cite{Furusaki95}. It also agrees perfectly with NRG results at large or small transmission (solid black lines, Fig.~\ref{fig:FL}) \cite{note:toulouse}.


\textit{Nonequilibrium transport.--} 
Our exact result for the conductance away from thermal equilibrium, Eq.~\ref{eq:FL}, is a universal function of $eV/T^{*}$ and $T/T^{*}$. 
Importantly, it is applicable whenever $T^* \ll T_K$, irrespective of  $T_K/D$. 
Unlike its NFL counterpart, the universal FL crossover \emph{can} therefore be observed in experiments performed at large transmissions -- a unique feature of \emph{charge}-Kondo devices.
With $T_K \sim E_c$ near perfect transmission, the FL crossover can realistically be explored over a sizable range of $T$ and $eV$, while still ensuring $T^{*}\ll T_K$.

Our predictions for the nonequilibrium quantum critical conductance along the FL crossover are summarized in Fig.~\ref{fig:noneq}, taking the experimental value $E_c$=$290$mK~\cite{Iftikhar15}.
When $\max \{ T,eV \} \gg T^{*}$, finite $\Delta E$ or $\Delta J$ give negligible corrections to the NFL conductance $G \sim \frac{e^2}{2h}$. But for $T \ll eV $, the RG flow is cut by the nonequilibrium scale $eV$. As shown in the main panel of Fig.~\ref{fig:noneq}, the conductance reduces to zero as $eV$ is reduced (the FL fixed point is reached when $eV \ll T^*$, see bottom right panel). By tuning the gates closer to criticality, $T^*$ reduces, causing the width of the anti-peak to reduce. However, this reduction is cut off when $eV \sim T$, as shown in the top right panel at the experimental base temperature $T$=$11.5$mK. 
These direct signatures of nonequilibrium criticality should therefore be  observable under the present experimental conditions~\cite{Iftikhar15}.


\textit{Concluding remarks.--}
Our analysis of the charge 2CK device provides a stringent test for universality and scaling in nanoscale devices, demonstrating the importance of being in the scaling regime to extract true universal results. We showed that the experiment of Ref.~\cite{Iftikhar15} precisely realizes the charge 2CK model, including both universal and nonuniversal properties. 
The device is a rare example in which the long sought \emph{nonequilibrium} structure of a NFL critical point can be explored experimentally, and compared with exact theoretical results.

Finally, we highlight a perspective on these results, connected with the ongoing search for Majorana fermions. 
The quantitative agreement between theory and experiment over 9 orders of magnitude in $T/T_K$  proves that this device realizes a non-Fermi liquid state involving a free Majorana localized on the dot, described by the 2CK critical fixed point. These results therefore \emph{unambiguously} establish the existence of Majorana fermions in this frustrated strongly interacting system.


\textit{Acknowledgments.--}
We thank M. Goldstein and F. Pierre for discussions. Experimental data from Ref.~\cite{Iftikhar15} reproduced with permission. AKM and LF acknowledge the D-ITP consortium, a program of the Netherlands Organisation for Scientific Research (NWO) that is funded by the Dutch Ministry of Education, Culture and Science (OCW). ES acknowledges funding by the Israel Science Foundation Grant No.~1243/13, and the Marie Curie CIG Grant No.~618188. 



\begin{thebibliography}{38}
\bibitem [{not({\natexlab{a}})}]{note:numop}%
  \BibitemOpen
  \bibinfo {note} {The total dot number operator is $\hat{N}_{\downarrow} =
  \sum_{\alpha k}c_{\alpha\downarrow k}^{\dagger}c_{\alpha \downarrow
  k}^{\phantom{\dagger}}$ such that $\hat{N}_{\downarrow}
  |N_{\downarrow}\rangle = N_{\downarrow}|N_{\downarrow}\rangle$. The charging
  Hamiltonian is $H_c=E_c ( \hat{N}_{\downarrow}-N_0 )^2$, with $ N_0$
  controlled by gate $V_g$.}\BibitemShut {Stop}%
\bibitem [{not({\natexlab{b}})}]{note:dos}%
  \BibitemOpen
  \bibinfo {note} {The detailed form of the electronic density of states could
  be neglected at the experimental base temperatures.}\BibitemShut {Stop}%
\bibitem [{not({\natexlab{c}})}]{note:nrg}%
  \BibitemOpen
  \bibinfo {note} {NRG data were obtained using discretization parameter
  $\Lambda=3$ and retaining $N_K=10000$ states per iteration, within symmetry
  setting $[\text{U(1)}_{\text{charge}}]^2\times\text{U(1)}_{\text{spin}}$.
  Results were $z$-averaged over $N_z=2$ calculations.}\BibitemShut {Stop}%
\end{thebibliography}
%


\end{document}